\documentclass[twocolumn,showpacs,preprintnumbers,amsmath,amssymb]{revtex4}

\usepackage{graphicx}
\usepackage{dcolumn}
\usepackage{bm}

\begin{document}

%
\newcommand{\bea}{\begin{eqnarray}}
\newcommand{\eea}{\end{eqnarray}}
\newcommand{\be}{\begin{equation}}
\newcommand{\ee}{\end{equation}}
%
\newcommand{\xbf}[1]{\mbox{\boldmath $ #1 $}}

\title{Non-spherical proton shape and hydrogen hyperfine 
splitting~\footnote{published in Can. J. Phys. {\bf 87}, 773 (2009).}}

\author{A. J. Buchmann}
\affiliation{
Institute for Theoretical Physics \\
University of T\"ubingen \\
D-72076 T\"ubingen, Germany}
\email{alfons.buchmann@uni-tuebingen.de}

\begin{abstract}
We show that the non-spherical charge distribution of the proton 
manifests itself in hydrogen hyperfine 
splitting as an increase (in absolute value) of the proton Zemach radius 
and polarization contributions.
\end{abstract}


\maketitle

\section{Introduction}
Hydrogen hyperfine splitting (hfs), which is predominantly 
due to the interaction of electron and proton magnetic moments, 
is an interesting observable.
It provides not only precision tests for quantum electrodynamics, 
but also valuable information on proton structure and strong interactions.
According to Fermi's theory~\cite{fer30}, the energy difference
between the two hydrogen hyperfine states is 
\be
\label{fermienergy_final}
E_F= \frac{8}{3}\,\frac{\alpha^3}{\pi} \, 
\frac{m_e^3 \, m_p^3 }{(m_e+m_p)^3}\, 
\mu_e \, \mu_p,
\ee
where $\alpha$ is the fine structure constant, $m_e$ is the mass of the 
electron and $\mu_e$ its magnetic moment; $m_p$ and $\mu_p$  are the 
proton mass and magnetic moment. Numerically, this gives 
$E_F= 1, 418, 840.11$ kHz using recent values 
for these fundamental constants~\cite{moh08}.

The interplay between experiment and theory has been particularly fruitful 
in the case of hydrogen hfs. For example, when the original 
measurement~\cite{naf47} exceeded the prediction of 
Eq.(\ref{fermienergy_final}) by 0.26 $\%$,
which was far more than the experimental accuracy at that time, 
this stimulated the first quantum electrodynamics (QED) 
calculation of the electron anomalous magnetic moment~\cite{sch47}. 
The latter was then identified as the major reason for the 
$\sim$ MHz discrepancy between experiment and Fermi's formula. 
Further studies revealed that hydrogen hyperfine splitting 
is not only affected by global 
proton properties such as its mass, charge, and magnetic moment, but is  
also sensitive to the details of the spatial charge and current distributions
in the proton and its excited states~\cite{zem56,dre67}. These nucleon 
structure effects contribute to hfs at the level of several tens of kHz.  

In the meantime, the transition frequency between the two hyperfine states 
has been measured~\cite{ess71} with an experimental uncertainty of about 
1 mHz 
\be
\label{hfs_exp}
E^{HFS}_{exp} = 1,420,405,751.7667 \pm 0.0009 \, {\rm [Hz]}, 
\ee
corresponding to a relative accuracy of $10^{-12}$. 
Thus, hydrogen ground state hfs is one of most precisely measured physical 
quantities. On the other hand, the accuracy of present calculations of  
nucleon structure effects in hydrogen hfs is at best of order $10^{-6}$ 
and hence many orders of magnitude lower than the precision with which the 
fundamental constants and QED corrections are known.  Therefore, hydrogen hfs 
can be used as a high precision probe for investigating fine details 
of proton structure.
 
Another way of studying the structure of the nucleon is electron-nucleon
scattering.  Employing polarized electron beams and hydrogen 
targets, it has recently become possible to experimentally determine the 
$p \to \Delta^+(1232)$ charge quadrupole transition form factor~\cite{ber07}.
It has been proposed that the quadrupole excitation 
of the nucleon $N(939)$ to the $\Delta$(1232) resonance is closely related to 
a quadrupole deformation of the nucleon's ground state charge distribution 
as reflected by a positive intrinsic quadrupole moment~\cite{Hen01} and 
an intrinsic charge quadrupole form factor~\cite{buc05}. 
This interpretation is based on relations between nucleon ground state 
and $N \to \Delta$ transition form factors that follow from  
broken strong interaction symmetries.

The purpose of this paper is to explore in which way and to what extent 
the proton's non-spherical charge density affects hydrogen hfs. 
As a spin 1/2 particle the proton does not have a spectroscopic 
quadrupole moment and its non-spherical charge distribution does not 
result in a quadrupole interaction term in the hfs energy shift 
formula~\cite{buc05}. Nevertheless, deviations from a spherically 
symmetric proton charge distribution are detectable in hydrogen hfs  
via the quadrupole excitation of the nucleon to the $\Delta^+(1232)$ 
resonance by the atomic electron (polarization shift) and 
via their effect on the electromagnetic size of the proton (Zemach radius).

The paper is organized as follows. In sect.~\ref{sec:structure} we discuss 
the electromagnetic $N\to \Delta$ transition and what we have learned from it
about the geometric shape of the nucleon in some detail. 
In sect.~\ref{sec:implications} the implications of the nucleon's 
non-sphericity  for the hydrogen atom hyperfine splitting are investigated. 
The paper closes with a summary of our results and an outlook.

\begin{figure*}[htb]
\centering{
\includegraphics{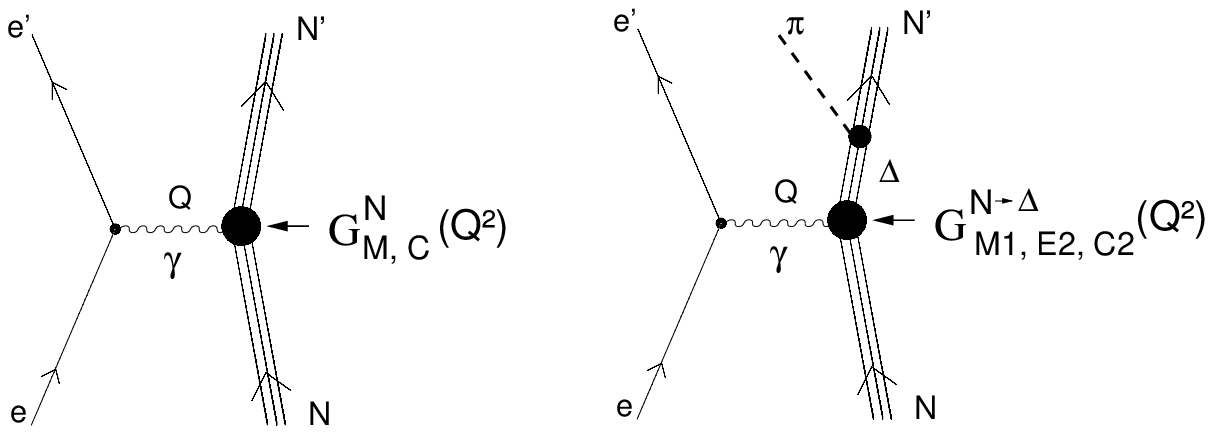}
\caption{\label{figure:scattering} 
Left: Probing of nucleon structure via elastic electron-nucleon
scattering $e\, N \to e' \, N'$ involving the exchange of a single 
virtual photon $\gamma$ of four-momentum $Q=(\nu, -{\bf q})$ 
with $\nu$ and ${\bf q}$ being its energy and three-momentum transfer. 
Nucleon structure information is contained in the magnetic dipole 
form factor $G^{N}_{M}(Q^2)$ and charge monopole form factor $G^{N}_{C}(Q^2)$.
Right: Inelastic electron-nucleon scattering 
$e \, N \to e' \, \Delta \to e' N' \pi$ (electro-pionproduction).
The electromagnetic excitation of the $\Delta(1232)$ resonance is described
by three transition form factors
$G^{N \to \Delta}_{M1}(Q^2)$, $G^{N \to \Delta}_{E2}(Q^2)$,
and $G^{N \to \Delta}_{C2}(Q^2)$.}
}
\end{figure*}

\section{Electromagnetic $N \to \Delta$ transition and nucleon shape}
\label{sec:structure}

\subsection{Elastic and ineleastic electron scattering}
\label{subsec:scattering}

Nucleon structure information is encoded 
in two elastic electromagnetic form factors, namely the charge monopole 
$G_{C}^N(Q^2)$ and magnetic dipole $G_{M}^N(Q^2)$ form factors
as indicated by the black dot in Fig.~\ref{figure:scattering}~(left). 
These form factors have been measured in elastic electron-proton and 
electron-deuteron (neutron) scattering 
experiments performed at various laboratories. In particular, it has  
been shown that the proton has a finite charge radius of about
0.9 fm~\cite{sim80}. 
In addition, the Fourier transforms of the elastic form factors 
have provided information on the radial variation of the charge
$\rho(r)$ and current ${\bf j}(r)$ densities 
of the proton~\cite{fri03}.

Inelastic electron-proton scattering with the production of a single 
pion (electro-pionproduction) has revealed that the proton has a 
rich spectrum of excited states~\cite{Bur94}. 
Its lowest lying excited state with spin 3/2 and isospin 3/2, 
the $\Delta(1232)$ resonance, plays a special role because it has 
the largest production cross section 
and its properties are most closely related to those of the nucleon 
ground state $N(939)$. Parity invariance of the electromagnetic interaction 
and angular momentum conservation restrict the $N \to \Delta$ excitation  
to magnetic dipole (M1), electric quadrupole (E2), and charge or Coulomb 
quadrupole (C2) transitions with corresponding
transition form factors as depicted by the large black dot 
in Fig.~\ref{figure:scattering}~(right). 
The nonzeroness of the C2 form factor indicates that 
the nucleon charge distribution is not spherically symmetric~\cite{ber07} 
but has an angular dependence $\rho({\bf r})=\rho(r, \Theta, \Phi)$.
In the following we review the connection between the quadrupole excitation 
of the $\Delta$(1232) resonance and nucleon ground state deformation 
using strong interaction symmetries as a guide. 

\subsection{Spin-flavor symmetry and electromagnetic form factor relations}
\label{subsec:symmetries}

Aside from SU(2) isospin and SU(3) flavor symmetries, strong interactions 
are also approximately invariant under the higher SU(6) spin-flavor
symmetry. The latter unites the spin 1/2 flavor octet baryons 
($2 \times 8$ states), among them the familiar proton and neutron, and the
spin 3/2 flavor decuplet baryons ($4 \times 10$ states), including
the four $\Delta(1232)$ states into a common {\bf 56}-dimensional mass 
degenerate supermultiplet~\cite{Gur64}. 
We now understand that the underlying field theory of strong interactions,
quantum chromodynamics (QCD), possesses a spin-flavor symmetry which
is exact in the large $N_c$ limit,
where $N_c$ denotes the number of colors. Moreover, for finite
$N_c$, spin-flavor symmetry breaking operators can be classified
according to the powers of $1/N_c$ associated with them.
This leads to a perturbative expansion scheme
for QCD processes that works at all energy scales~\cite{Wit79}.

\begin{figure}[htb]
\centering{
\resizebox{0.45\textwidth}{!}
{\includegraphics{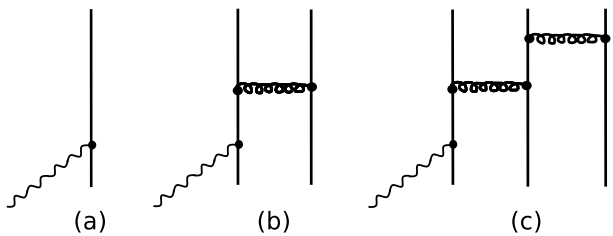}}
\caption{
\label{figure:currents} 
Fundamental photon-quark processes contributing 
to the form factors in Fig.~\ref{figure:scattering}:
(a) one-quark current ($\rho_{[1]},{\bf j}_{[1]})$, 
(b) two-quark gluon exchange current $(\rho_{[2]}, {\bf j}_{[2]})$,
(c) three-quark gluon exchange current $(\rho_{[3]}, {\bf j}_{[3]})$.}}
\end{figure}

For $N_c=3$ we may just as well employ a parametrization method~\cite{Mor89}, 
which incorporates SU(6) symmetry and its breaking similar to the $1/N_c$ 
expansion. The basic idea is to write for the observable under investigation 
the most general spin-flavor operator.
Generally, this is a sum of one-, two-, and three-quark
operators in spin-flavor space multiplied by {\it a priori} unknown constants
which parameterize the orbital and color space
matrix elements, and which are determined from experiment.
A multipole expansion of the nucleon charge density operator $\rho$ 
in spin-flavor space up to quadrupole terms leads to the following invariants 
\begin{eqnarray}
\label{structure}
\rho  \! \! \! \! &  =  & \! \! \! \! \rho_{[1]} + \rho_{[2]} + \rho_{[3]} =
A \sum_i^{3} e_i \, {\bf 1} -  \! 
\left ( \! B \sum_{i \ne j}^{3} e_i\! 
+ \!C \!\! \sum_{i\ne j \ne k}^3 \! \! e_k \! \! \right )\!\! \nonumber \\
& & \biggl [ 
2 \! \! \underbrace{\bm{\sigma}_i \cdot \bm{\sigma}_j}_{{\rm spin \ scalar}} 
\! \!
- \underbrace{( 3 \bm{\sigma}_{i\, z} \, \bm{\sigma}_{j \, z} 
-\bm{\sigma}_i \cdot \bm{\sigma}_j )}_{{\rm spin \ tensor}} 
\biggr ], 
\end{eqnarray}
where $\bm{\sigma}_{i \, z}$ is the $z$-component of the Pauli spin
matrix of quark $i$, and $e_i=\frac{1}{6}(1 + 3\, \bm{\tau}_{i \, z})$ 
is the quark charge where $\bm{\tau}_{i \, z}$ is the third component
of the Pauli isospin matrix. 
The constants $A$, $B$, and $C$ contained in the one-, two- 
and three-quark charge density 
operators parametrize the orbital and color matrix elements
so that $\rho$ is only an operator in spin-flavor space.
The factors 2 and -1 in front of the spin scalar and spin tensor terms 
are dictated by group theory and reflect the fact that both terms
arise from a common SU(6) spin-flavor tensor~\cite{buc05}.
In terms of fundamental processes (see Fig.~\ref{figure:currents}), 
the one-quark operator in Eq.(\ref{structure}) represents valence quarks 
whereas the two-, and three-quark operators provide 
an effective description of the quark-antiquark 
degrees of freedom in the $N$ and $\Delta$.
For the two-body current in Fig.~\ref{figure:currents}(b)  
this is explained in more detail in Ref.~\cite{Buc89}.

Evaluating the charge operator in Eq.(\ref{structure}) between 
spin-flavor wave functions~\cite{Lic78} of the ${\bf 56}$ dimensional SU(6) 
ground state multiplet, in particular, for the neutron and between the 
initial proton and final $\Delta^+$ states, leads to
\bea
\label{su6calc}
r_n^2 & =& \langle 56_{ \, n \,} \vert \rho_{[2]} + \rho_{[3]} 
\vert 56_{n\,} \rangle 
= 4 (B - 2 C), \nonumber \\
Q_{p \to \Delta^+} & = &  
\langle 56_{\Delta^+} \vert \rho_{[2]} + \rho_{[3]} 
\vert 56_p \rangle = 
2 \sqrt{2} (B-2C).  
\eea
Note that in this approach, valence quarks make no contribution
and both observables are governed by quark-antiquark degrees of freedom 
in the nucleon. Hence, the following relation between the 
transition quadrupole moment $Q_{p \to \Delta^+}$
and the neutron charge radius $r^2_n$ is obtained
\be
\label{rnquad}
 Q_{p \to \Delta^+}=  
\frac{1}{\sqrt{2}} \, r^2_{n}.  
\ee
This relation was originally derived in the constituent quark model 
with two-quark exchange currents~\cite{Buc97} and shown to hold after
including three-quark operators~\cite{Hes02}.
It was found that Eq.(\ref{rnquad}) is the zero momentum transfer 
limit of a more general relation~\cite{buc04} between the $N \to \Delta$ 
charge quadrupole transition form factor  $G^{N\to \Delta}_{C2}(Q^2)$ and the 
elastic neutron charge form factor $G_C^n(Q^2)$
\bea
\label{ffrel2}
G_{C2}^{N \to \Delta}(Q^2) & = & - \frac{3 \sqrt{2}}{Q^2} \, G_{C}^{n}(Q^2), 
\nonumber \\
Q_{N\to \Delta} :\, =G_{C2}^{N \to \Delta}(0) & = &  
\frac{1}{\sqrt{2}} \, r^2_{n}, \nonumber \\ 
r^2_{C2} & = &  \frac{7}{10}\, \frac{r^4_n}{r^2_n},
\eea
which is valid for both the $p \to \Delta^+$ and $n \to \Delta^0$ 
transitions, and that the quadrupole transition radius is determined 
by the fourth and second radial moments of the neutron charge distribution.

In addition, SU(6) spin-flavor symmetry leads to the following
relations~\cite{Gur64} between the neutron ground state and the $N \to \Delta$ 
magnetic form factors $G_{M1}^{N \to \Delta}(Q^2)  =  -\sqrt{2}\, G_M^n(Q^2)$, 
and at $Q^2=0$ between the neutron and transition magnetic moments 
$\mu_{N \to \Delta}= -\sqrt{2} \, \mu_{n}$. 
With the help of Eq.(\ref{ffrel2}) and the magnetic form factor relations,
the C2/M1 ratio in electromagnetic $\Delta(1232)$ excitation can be expressed 
in terms of the neutron elastic form factors as follows~\cite{buc04} 
\bea
\label{ffrel3}
\frac{C2}{M1}(Q^2) &:= & \frac{\vert {\bf q} \vert \, m_N}{6}  
\, \frac{G_{C2}^{N \to \Delta}(Q^2)}{G_{M1}^{N \to \Delta}(Q^2)}= 
\frac{ \vert {\bf q}\vert m_N}{2 \,Q^2} \, 
\frac{G_{C}^{n}(Q^2)}{G_M^n(Q^2)}, \nonumber \\
\frac{C2}{M1}(0) & = &  \frac{(m_{\Delta}-m_N)\, m_N}{12} \, 
\frac{r_n^2}{\mu_n}, 
\eea
where $\vert {\bf q} \vert$ is the modulus of the photon's three-momentum 
and where $m_N$, $m_{\Delta}$ are the nucleon and $\Delta$ 
masses. The theoretical uncertainty of this relation is mainly 
due to third order SU(6) symmetry breaking operators of order $1/N_c^2$  
(three-quark currents) violating the magnetic form factor 
relation. It is estimated that such correction terms could lead to 
$10\%$ decrease of $\vert C2/M1\vert$.

\subsection{Comparison with experiment}
\label{subsec:comparison}

\begin{figure*}[t]
\includegraphics[height=0.35\textheight]{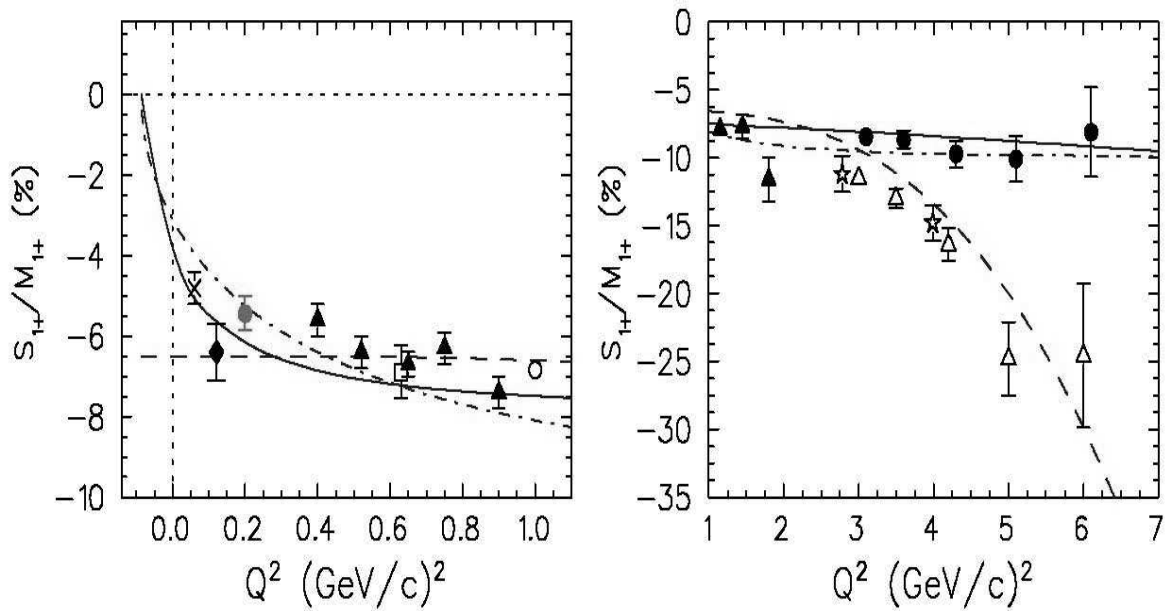}
\caption{
\label{C2_m1maid07}
The $C2/M1(Q^2)\equiv S_{1+}/M_{1+}(Q^2)$ ratio for 
low (left) and high (right) four-momentum transfers. 
The full curve is a fit of the experimental $C2/M1$ ratio as determined 
from the world electro-pionproduction data.
The dashed-dotted curve is calculated using the form factor relation of 
Eq.(\ref{ffrel3}). The open triangles and
the dashed curve are from a previous analysis of the same data~\cite{ung06}.
Figure taken from Ref.~\cite{dre07}. }
\end{figure*}

In Fig.~\ref{C2_m1maid07}, the experimental $C2/M1$ ratio as measured in 
electro-pionproduction  is shown. The full curve represents the Maid 2007 
analysis~\cite{dre07} 
of the world $C2/M1$ data, while the dashed-dotted line is based on 
Eq.(\ref{ffrel3}) which relates the inelastic $N\to\Delta$ and the 
elastic neutron form factors. For definiteness we use a 
Galster parametrization~\cite{Gal71} 
\bea
\label{galster}
G_C^n(Q^2) & = & -a \frac{\tau}{1+d \tau} \, G_M^n(Q^2), \quad
\tau= \frac{Q^2}{4 m_N^2}, \nonumber \\ 
G_M^n(Q^2) & = & \mu_n \left ( 1+\frac{Q^2}{\Lambda_M^2} \right)^{-2} 
\eea
for the experimental neutron charge form factor, where 
$G_M^n(Q^2)$ is a dipole representation of the neutron magnetic form factor
with $\Lambda_M$ being the dipole mass.
The parameters $a$ and $d$ are related to the second and fourth moment 
of the neutron charge distribution respectively~\cite{buc04} and have
the numerical values $a=0.9$ and $d=1.75$. 
As is clear from Fig.~\ref{C2_m1maid07}~(left), our theory 
agrees well with the $C2/M1$ data at low momentum transfers. 
In particular, at 
the real photon point $Q^2= 0$ we get from Eq.(\ref{ffrel3}) using the 
experimental neutron charge radius and magnetic moment $C2/M1=-0.035$, 
which is in good agreement with determinations of this ratio 
based on the experimental $E2/M1$ ratio measured in photo-pionproduction 
and Siegert's theorem relating $E2$ and $C2$ 
form factors~\cite{Buc97}.

Moreover, the transition quadrupole moment has been extracted from 
photo- and electro-pion\-pro\-duction data as
$Q_{p \to \Delta^+}({\rm exp})=-0.108(9)$ fm$^2$~\cite{Bla01} and 
$Q_{p \to \Delta^+}({\rm exp})=-0.0846(33)$ fm$^2$~\cite{Tia03}, 
in good agreement with the prediction 
$Q_{p \to \Delta^+}({\rm theory})=-0.0820(20)$ fm$^2$
based on Eq.(\ref{rnquad}) and the experimental neutron 
charge radius $r_n^2=-0.1161(22)$ fm$^2$.
Concerning the transition quadrupole radius, 
we obtain from  Eq.(\ref{ffrel2}) and experimental 
values for the radial moments~\cite{buc04} of the neutron charge distribution
$r_{C2} = 1.43 \, {\rm fm}$. 
The approximate equality $r_{C2} \approx r_{\pi}$, 
where $r_{\pi}$ is the pion Compton wavelength suggests that $r_{C2}$
measures the spatial extension of the $q{\bar q}$ pair distribution 
in the nucleon. It would be interesting to determine this radius 
experimentally. 

At higher momentum transfers shown in Fig.~\ref{C2_m1maid07}~(right) 
the extraction of individual electromagnetic multipoles from
the raw cross section data is more difficult as is 
evident from the difference between 
the Jefferson Lab~\cite{ung06} and Maid 2007~\cite{dre07} analyses
of the same raw data.
The latter analysis, which leads to smaller $\vert {\rm C2/M1}\vert$ 
(filled circles) 
than the former (open triangles) is in much better agreement with our theory
(dashed-dotted curve).
Extrapolating Eq.(\ref{ffrel3}) to $Q^2\to \infty$ we find~\cite{buc04}
\be
C2/M1(Q^2\to \infty) =-\frac{1}{4} \frac{m_N}{m_{\Delta}}\, 
\left ( \frac{a}{d} \right ). 
\ee
consistent with perturbative QCD,
which states that $C2/M1$ asymptotically 
approaches a small negative constant.

\subsection{Intrinsic quadrupole form factor of the nucleon}
\label{subsec:intrinsicqm}

For an interpretation of these results it is important to distinguish
between the {\it spectroscopic} and {\it intrinsic} quadrupole moment of 
a particle. 
It is known that a vanishing spectroscopic quadrupole moment does not 
necessarily imply a spherically symmetric charge distribution. 
For deformed spin 0 and spin 1/2 nuclei this has led to the general concept 
of an intrinsic quadrupole moment, which can be defined for different nuclear 
models.  The notion of an intrinsic quadrupole moment allows to interpret 
measurable quantities such as transition quadrupole moments in terms of the 
geometric shape of the ground state. 

The geometric shape of a spatially extended particle is determined by 
its {\it intrinsic} quadrupole moment,
\be 
Q_0=\int \! \!d^3r \, \rho({\bf r}) \,  (3 z^2 - r^2), 
\ee
which is defined with respect to the body-fixed frame. 
If the charge density is concentrated along the $z$-direction 
(symmetry axis of the particle),  
the term proportional to $3z^2$ dominates, $Q_0$ 
is positive, and the particle is prolate (cigar-shaped).
If the charge density is concentrated in the equatorial plane perpendicular
to $z$, the term proportional to $r^2$ prevails, $Q_0$
is negative, and the particle is oblate (pancake-shaped).

We calculated the intrinsic quadrupole moment of the proton and 
$\Delta^+$ in the quark model including two-body exchange 
currents~\cite{Hen01}, which effectively describe the 
quark-antiquark degrees of freedom in the nucleon, and found 
\be
\label{intquad}
Q_0^p = -r^2_n, \qquad Q_0^{\Delta^+}  =  r^2_n. 
\ee
Thus, the intrinsic quadrupole moment of the proton is given
by the negative of the neutron charge radius and is therefore
positive, whereas the intrinsic quadrupole moment of the $\Delta^+$ 
is negative. This corresponds to a prolate proton and an oblate
$\Delta^+$ shape. 
The model results also suggest that the nonsphericity of the 
proton charge density is mainly connected with 
collective quark-antiquark degrees of freedom, the distribution of
which has a prolate shape. 

The concept of an intrinsic quadrupole moment of the nucleon can
be generalized to an intrinsic quadrupole charge distribution and a
corresponding form factor~\cite{buc05}.
To show this, we first decompose the proton and neutron charge form factors
in two terms $G_{sym}$ and $G_{def}$, 
coming from the spherically symmetric and the intrinsic quadrupole 
part of the physical charge density repectively
\bea
\label{voldefdecomp}
G_C^p(Q^2) \!& = & \!G_{sym}^p(Q^2) -\frac{1}{6} \, Q^2 \, G_{def}(Q^2),
\nonumber \\
G_C^n(Q^2)\! & = & \! G_{sym}^n(Q^2) + \frac{1}{6} \, Q^2 \, G_{def}(Q^2).
\eea
The factor $Q^2$ in front of $G_{def}$ arises for dimensional reasons
and guarantees that the normalization of the charge form factors
is preserved. In coordinate space this corresponds to the usual multipole
decomposition of the charge density
\be
\rho({\bf r})=\underbrace{\rho_0(r) Y^0_0({\bf r})}_{{\rm monopole}}
+ \underbrace{\rho_2(r)Y^2_0({\bf r})}_{{\rm quadrupole}} + \ldots,
\ee
where the $\rho_0(r)$ part gives rise to $G_{sym}(Q^2)$ and the $\rho_2(r)$ 
part is connected with $G_{def}(Q^2)$. In terms of fundamental
photon-quark processes $G_{def}$ arises mainly from two- and 
three-quark currents.

For the intrinsic charge quadrupole form factor
$G_{def}(Q^2)$ we find 
\bea
\label{intrinsicC2ff}
G_{def}(Q^2) & = & -\sqrt{2} \, G_{C2}^{N \to \Delta}(Q^2)
= \frac{6}{Q^2} \, G_{C}^n(Q^2), \nonumber \\ 
G_{def}(0) & = & -r_n^2=Q_0^p
\eea
where we have used Eq.(\ref{ffrel2}). 
The zero momentum limit follows from l' Hospital's rule 
and Eq.(\ref{intquad}).
This shows that $G_{def}(Q^2)$ as defined in 
Eq.(\ref{intrinsicC2ff}) is the proper generalization of the
intrinsic quadrupole moment $Q_0^p$ to finite momentum transfers.

To exhibit the effect of the intrinsic quadrupole form factor on the elastic
nucleon form factors we insert Eq.(\ref{intrinsicC2ff}) 
into Eq.(\ref{voldefdecomp}) and obtain
\bea
\label{isoscalar}
G_{C}^p(Q^2)  &=&  G_{sym}^p(Q^2)-G_{C}^n(Q^2) = 
\underbrace{G_C^{IS}(Q^2)}_{{\rm spherical}} - 
\underbrace{G_C^n(Q^2)}_{{\rm deformed}}, \nonumber \\
G_C^n(Q^2) & = & \frac{1}{6} \, Q^2 \, G_{def}(Q^2),
\eea
where the isoscalar nucleon charge form factor is defined as
$G_C^{IS}(Q^2)= G_C^p(Q^2)+G_C^n(Q^2)$ and $G_{sym}^n=0$.
Thus, the relation between the $N \to \Delta$ and neutron charge form
factors of section~\ref{subsec:symmetries} 
is seen here to have an important
implication for the nucleon itself, which can be summarized as:
{\it The neutron charge form factor is an
observable manifestation and quantitative measure of
the nucleon's intrinsic quadrupole form factor. The latter
manifests itself also in the proton charge form factor.}

There are several observable consequences of Eq.(\ref{intrinsicC2ff}) and 
Eq.(\ref{isoscalar}) as discussed in Ref.~\cite{buc05}. 
At low $Q^2$ the nucleons's prolate deformation is 
reflected in a proton charge radius increase by an amount $-r_n^2$,
or more directly by a newly introduced size parameter 
$r_{def}^2=r^2_{C2}=(7/10) (r_n^4/r_n^2)$ that can be 
experimentally determined.  
At intermediate $Q^2$ it leads to the conclusion that 
the dip structure observed in the proton
charge form factor~\cite{fri03} at around $Q^2 \approx 0.2$ GeV$^2$
is due to a corresponding structure in the neutron charge form factor
at the same $Q^2$.
Finally, at high $Q^2$ it 
explains the observed decrease of the charge over magnetic
form factor ratio~\cite{pun05}. 

We close this section by stating that our introduction of an intrinsic 
quadrupole moment and quadrupole form factor of the nucleon should be viewed 
as an attempt to explore the consequences of the experimental
sign and size of the $N\to \Delta$ quadrupole transition form factor 
for nucleon ground state structure, which in turn has a bearing 
on hydrogen hyperfine splitting.


\section{Hydrogen hyperfine splitting and non-spherical proton shape}
\label{sec:implications}

\subsection{Fermi energy}

It is well known that the ground state energy of atomic hydrogen is 
split into two levels due to the interaction of the magnetic moments
of the electron $\bm{\mu}_e$ and proton $\bm{\mu}_p$, 
which can be either aligned ($F=0$) or antialigned ($F=1$).
The energy difference between these two states is of order 
 $10^{-6}$ eV which is small compared to the -13.6 eV binding energy of the 
ground state.
For spherically symmetric electronic S states, the magnetic dipole-dipole 
interaction Hamiltonian can be expressed as
\bea
\label{fermienergy} 
H_F & \!\!\!= \!\!\! & -\frac{2}{3} \, \bm{\mu}_p \cdot  \bm{\mu}_e \, 
\delta^{(3)}({\bf r}_p -{\bf r}_e) \nonumber \\ 
& \!\!\! = \!\! \!& 
\frac{2}{3} (1 + \kappa) \! \! \left (\frac{e}{2m_p}\right )\! \! \! 
\left (\frac{e}{2m_e} \right )  
\bm{\sigma}_p \cdot  \bm{\sigma}_e  
\delta^{(3)}({\bf r}_p\!-\!{\bf r}_e).
\eea
where ${\bf r}_p$ and ${\bf r}_e$ are the proton and electron
position coordinates and $\delta^{(3)}$ is 
the three-dimensional Dirac $\delta$ function.
The second equation follows after rewriting the magnetic moments 
in terms of spin operators  as 
$\bm{\mu}_p=(1+\kappa)\, [e/(2 m_p)] \, \bm{\sigma}_p$ 
and $\bm{\mu}_e=(1+a)\, [e/(2 m_e)] \, \bm{\sigma}_e$. 
By convention the proton anomalous magnetic moment 
$1+\kappa$ is included in the Fermi energy, while the anomalous 
electron magnetic moment $1+a$ is included in the QED corrections 
introduced in sect.~\ref{subsec:corrections}.

After taking matrix elements of Eq.(\ref{fermienergy})
between hydrogen ground state electron wave functions $\Psi_e({\bf r}_e)$ 
one obtains for the hyperfine level splitting
\bea 
\label{fermi_matrixlement}
E_F  = 
\frac{2}{3} \, (1 + \kappa) \, \left (\frac{e}{2m_p} \right )\, 
\left ( \frac{e}{2m_e} \right ) & & \! \! \! \! \!
\biggl (\langle \bm{\sigma}_p \cdot  \bm{\sigma}_e  \rangle_{F=1} \nonumber \\
-\langle \bm{\sigma}_p \cdot  \bm{\sigma}_e  \rangle_{F=0} \biggr ) 
\vert \Psi_e(0) \vert^2,
\eea
where $\delta^{(3)}({\bf r}_p -{\bf r}_e)$ 
has been evaluated 
for ${\bf r}_e={\bf r}_p= {\bf 0}$ (point nucleon limit). 
We use the standard normalization $N$   
of the hydrogen atom ground state wave function 
\be
\label{wf}
\vert \Psi_e(0) \vert^2 =
\left (\frac{1}{\sqrt{\pi}}\, \frac{1}{a_B^{3/2}} \right)^2= N^2,
\qquad a_B=\frac{1}{m_r \, \alpha},
\ee
where $a_B$ is the Bohr radius, which is given in 
terms of the reduced mass $m_{r}=m_p\, m_e/ (m_p+m_e)$ 
of the electron-proton system and the fine structure constant 
$\alpha=e^2/(4 \pi)$.
The spin matrix element in Eq.(\ref{fermi_matrixlement}) 
gives $ <\bm{\sigma}_p \cdot  \bm{\sigma}_e>_{F=1} =1$ and
$ <\bm{\sigma}_p \cdot  \bm{\sigma}_e>_{F=0} =-3$ so that 
one obtains the Fermi energy formula of Eq.(\ref{fermienergy_final}) and
the numerical value $E_F= 1, 418, 840.11$ kHz.

\subsection{QED and proton structure corrections to the Fermi energy}
\label{subsec:corrections}

It has become customary to express QED and proton structure corrections 
as parts per million of the Fermi energy, i.e., 
1 ppm =$10^{-6} E_F$=1.41884 kHz.
The most important corrections to the pointlike dipole-dipole 
interaction energy in Eq.(\ref{fermienergy_final}) are due to
(i) QED, (ii) nucleon recoil, (iii) finite proton size, and 
(iv) proton polarization effects~\cite{Dup03}.
First, there is the anomalous magnetic moment of the electron, 
which is mainly caused by the QED vertex correction. 
This and other smaller QED contributions~\cite{eid01} lead to a
1136.1 ppm increase of the theoretical hfs
\be 
\label{fermienergy_qed}
E_{QED}^{HFS} = E_F (1 + \delta_{QED}) =1, 420, 452.04 \, {\rm kHz} 
\ee
to be compared with the experimental value in Eq.(\ref{hfs_exp}).
One readily notices that there is still a discrepancy between theory
and experiment, namely $E_{QED}^{HFS}-E_{exp}^{HFS}=
46.29 \, {\rm kHz}$ or $ 32.63 \, {\rm ppm}$.

\begin{figure*}[tbh]
\centering{
\vspace{-1.5 cm} 
\includegraphics[height=0.45\textheight]{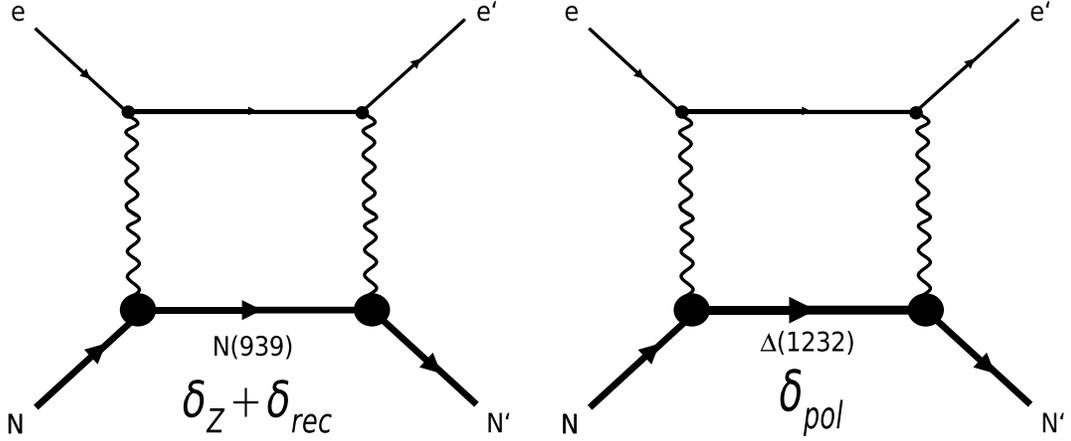}
\vspace{-1.5 cm} 
\caption{\label{figure:twophoton} 
Two-photon exchange diagrams (crossed photon diagrams 
are not shown) from which the elastic (Zemach and recoil) and inelastic 
(polarization) nucleon structure corrections to hydrogen hyperfine splitting 
are derived. The large black circles represent elastic (left) and
inelastic (right) nucleon electromagnetic form factors as 
measured in electron-nucleon scattering depicted 
in Fig.~\ref{figure:scattering}.
}}
\end{figure*}

Second, adding nucleon structure dependent relativistic recoil 
corrections $\delta_{rec}=5.85 \, {\rm ppm}$ to the theory~\cite{car08} this
discrepancy increases to $38.48$ ppm, i.e., a significant deviation
by which the theoretical value exceeds the measured one. 
Third, as a consequence of the proton's finite size, 
its magnetic moment is distributed over an extended 
spatial region. This weakens the magnetic interaction with the atomic electron 
and reduces the hfs. A first estimate of the proton size effect can be obtained
from an expansion of the electron wave function for small radial 
distances $\Psi_e(r) = N \exp(-r/a_B) = N \, ( 1 -r/a_B +\dots )$. 
The spatial extension of the proton magnetic moment distribution 
is then taken into account by evaluating the electron wave function
for $r_e=r_m\ne0$, where $r_m$ is the proton magnetic radius. 
This provides a correction term to the Fermi energy of the form~\cite{fen65}
\be
\label{fermienergy_finitesize}
E_{proton \, size}^{HFS} = E_F \left ( 1- 2 \,\frac{r_m}{a_B} \right )
\ee
and the numerical estimate of the proton size effect
$-2 \, r_m/a_B \approx -2 \cdot 10^{-5} \AA/0.5 \AA =-40$ ppm.
When these corrections are added to Eq.(\ref{fermienergy_qed}), one obtains
a reduction of the theoretical hfs which is of the right size 
to achieve agreement between theory and the experimental value in 
Eq.(\ref{hfs_exp}) at the ppm level. 
The recoil and finite size corrections
originate both from second order 
elastic electron-nucleon scattering depicted by the two-photon exchange 
diagram in Fig.~\ref{figure:twophoton}~(left) and  
are conventionally denoted as $\delta_{rec}$ and  $\delta_Z$. 

Fourth, there are also inelastic contributions (nucleon polarization),
which involve intermediate excited proton states, e.g., the $\Delta$(1232) 
resonance as shown in Fig.~\ref{figure:twophoton}~(right). 
The elastic and inelastic second order electron-proton interaction terms 
are generically referred to as proton structure contributions. 
The most important corrections to the Fermi energy are then
\be
\label{corrections}
E_{theory}^{HFS} = E_F \left ( 1+ \delta_{QED}  
+ \delta_{rec} + \delta_{Z} 
+ \delta_{pol} \right ).
\ee

\subsection{Zemach radius and proton shape}
\label{sec:zemach}

A careful analysis of the proton finite size correction
in hydrogen hyperfine splitting was performed  
by Zemach~\cite{zem56}. Assuming rigid (unpolarizable), 
spherically symmetric charge and magnetization distributions 
for the proton, the following expressions were derived
\bea
\label{zemach_def}
r_Z & = & 
-\frac{4}{\pi} \int_0^{\infty} \frac{dQ}{Q^2} \left \lbrack 
\, G_C^p(Q^2) \, \frac{G_M^p(Q^2)}{\mu_p} \, - 1 \right \rbrack, \nonumber \\
\delta_Z & = & -2 \, r_Z/a_B 
\eea 
where $G_C^p(Q^2)$ and $G_M^p(Q^2)$ are the elastic charge and
magnetic form factors of the proton. In contrast to the estimate 
in Eq.(\ref{fermienergy_finitesize}) the Zemach correction $\delta_Z$ 
depends on the details of {\it both}, charge and magnetic moment 
distributions.  The term $-1$ in the integrand of Eq.(\ref{zemach_def}) is 
necessary because the point nucleon limit, $G_C^p(0)=G_M^p(0)/\mu_p=1$, 
is already included in the Fermi energy $E_F$ and must be subtracted 
to avoid double counting. There is also a radiative correction term 
$\delta_{rad}$ due to electronic vacuum polarization~\cite{kar97} 
which changes $\delta_Z \to \delta_Z \, (1+ \delta_{rad})$.

In order to separate the effect of proton's non-spherical charge distribution
on the Zemach radius
we insert for $G_C^p(Q^2)$ the decomposition of Eq.(\ref{isoscalar}) into
Eq.(\ref{zemach_def}) and obtain
\bea
\label{zemach_decomp}
r_Z   =   
-\frac{4}{\pi} \int_0^{\infty} \! \frac{dQ}{Q^2} \! \! \!& &  \! \!  
\! \!   
\Biggl \lbrack \, \underbrace{G_C^{IS}(Q^2) \, 
\frac{G_M^p(Q^2)}{\mu_p} \, - 1}_{\rm spherical} \nonumber \\ 
& & -\underbrace{G_C^{n}(Q^2) \, \, \frac{G_M^p(Q^2)}{\mu_p}}_{\rm deformed} 
\Biggr \rbrack  \nonumber \\  
&  & \hspace{-2.5cm} =    r_Z(sym) + r_Z(def). 
\eea
Thus, the proton Zemach radius is decomposed into two terms 
coming from the spherically symmetric 
and non-spherical parts of the proton charge distribution respectively.
 
For an estimate of the spherically symmetric contribution 
we assume dipole forms for the isoscalar charge and proton 
magnetic form factors and obtain the compact two-parameter formula
\be
\label{two_dipoles}
r_Z(sym) = 
\frac{
3 \Lambda_I^4 
+9 \Lambda_I^3\,\Lambda_M 
+11 \Lambda_I^2\,\Lambda_M^2 
+9 \Lambda_I\,\Lambda_M^3 
+3 \Lambda_M^4}
{\Lambda_{I} \Lambda_M \, (\Lambda_{I} +\Lambda_M)^3 },
\ee
which hitherto has not appeared in the literature.
Here, $\Lambda_I$ and $\Lambda_M$ are the inverse size parameters  
of the isoscalar charge and proton magnetic form factors.
They are related to the corresponding mean square isoscalar charge 
and proton magnetic radii as $\Lambda^2_I=12/r^2_I$ with 
$r^2_I=r_C^2(p)+r_C^2(n)$ and $\Lambda^2_M=12/r_M^2(p)$.
In the limit, $\Lambda_I=\Lambda_M = \Lambda$, 
Eq.(\ref{two_dipoles}) reduces to $r_Z(sym)=35/(8 \, \Lambda)$,
i.e., a standard expression that has been used
by several authors~\cite{kar97,bod88,ver65}.
The contribution of the non-spherical part of the proton charge distribution
can also be analytically calculated if one uses 
the two-parameter Galster form of Eq.(\ref{galster}) for $G_C^n(Q^2)$ 
\bea
\label{Zemach_def}
r_Z(def) & \! \! \! = \! \! \! & -\frac{1}{6}\, r_n^2 (\Lambda_M \, m) \, 
\nonumber \\
& & \frac{
16 \Lambda_M^3
+29 \Lambda_M^2 \, m 
+20 \Lambda_M \, m^2 
+5 m^3 }
{8\, (\Lambda_M  +m)^4}, 
\eea
where we have used the same dipole mass $\Lambda_M$ for the proton
and neutron magnetic form factors.
The quantities $r_n^2$ and $m^2$ are related to the 
Galster parameters $a$ and $d$ as  
$r_n^2=3a\mu_n/(2 m_N^2)$ and $m^2 := 4m_N^2/d$. Eq.(\ref{two_dipoles}) 
and Eq.(\ref{Zemach_def}) are two main results of this paper.

Inserting experimental values for the dipole parameters $\Lambda_M$ and
$\Lambda_I$, as well as for $r_n^2$ and the Galster parameter $d=1.75$, 
one finds the numerical result
\bea
\label{zemach_num}
r_Z  & = & r_Z(sym) + r_Z(def)  =  1.0627 \ {\rm fm} + 0.0464 \ {\rm fm} 
\nonumber \\ 
& = & 1.1091 \ {\rm fm}
\eea
corresponding to $\delta_Z=-(40.14+1.75)$ ppm $=-41.89$ ppm.
The proton deformation contribution is numerically of the same size but 
of opposite sign as the neutron Zemach radius~\cite{fri05}. The sign change 
is obvious from the definition in Eq.(\ref{zemach_decomp}) and 
the approximate equality in magnitude $r_Z(def) \approx \vert r_Z(n)\vert$ 
follows from the near equality of the normalized proton and neutron magnetic 
form factors in the relevant moment transfer range. 
Including the radiative correction gives our final result for the 
proton Zemach 
radius $\delta_Z=-41.89$ ppm $(1+0.0151) =-42.52$ ppm. 
With the Zemach radius contribution included one finds 
that the discrepancy between theory and experiment reduces to 
$(38.48-42.52)\, {\rm ppm} = -4.04 \, {\rm ppm}$.

To obtain a better estimate of the effect of the 
proton's non-spherical shape on $r_Z$ one would have to determine the proton 
and neutron charge form factors more accurately in the low momentum transfer 
region where possible deviations from the smooth dipole and
Galster fitting curves may affect $r_Z(def)$ significantly.
Electron-proton scattering experiments dedicated to explore this 
low momentum transfer region with higher precision are being 
planned~\cite{arr07}.


\subsection{Proton quadrupole polarization shift}
\label{sec:polarization}

The proton polarization shift $\delta_{pol}$ in hydrogen hfs is caused by 
two-photon exchange diagrams with nucleon resonances as intermediate states,
of which the lowest-lying $\Delta(1232)$ as shown 
in Fig.~\ref{figure:twophoton} (right) is expected to be most important.  
Here, we focus on those diagrams where one of the photons
is a longitudinal charge quadrupole (C2) photon that probes the non-spherical 
charge distribution in both the ground and excited states. 

The polarization shift is usually defined in terms of integrals
over the two spin-dependent structure functions $g_1(x,Q^2)$ 
and $g_2(x,Q^2)$ of the proton, where $x=Q^2/(2 m_p \nu)$ is the 
Bjorken scaling variable and $\nu$ is the energy transfer carried by the 
virtual photon (see Fig.~\ref{figure:scattering}). 
The following formulae for proton polarization in hydrogen hfs 
have been established~\cite{car08}  
\bea
\label{polarization}
\delta_{pol} & \! \! \!= \! \! \!& 
\frac{\alpha \, m_e}{2 \pi \, m_p (1+ \kappa)} \left ( \delta_1 + \delta_2
\right ), \nonumber \\
\delta_1 & \! \! \! = \! \! \!  &
\frac{9}{4} \int_0^{\infty} \frac{dQ^2}{Q^2} 
\Biggl \lbrack F_2^2(Q^2) + \nonumber \\
& & \frac{8 m_p^2}{Q^2} 
\int_0^{x_{th}} \! \! \! dx \, \beta_1(\eta) \, 
\, g_1(x, Q^2) \Biggr \rbrack,  \nonumber \\
\delta_2 & \! \! \! = \! \! \! & 
-24 m_p^2 \int_0^{\infty} \! \! \frac{dQ^2}{Q^4} 
\left \lbrack \int_0^{x_{th}} \! \! \! dx \, \beta_2(\eta) \,  
\, g_2(x, Q^2) \right \rbrack, 
\eea
where $x_{th}= Q^2/(2 m_p \,m_{\pi} +m_{\pi}^2 +Q^2)$ is the threshold
for one-pion production with $m_{\pi}$ being the pion mass,
and the functions $\beta_1(\eta)$ and $\beta_2(\eta)$ are defined as
\bea
\beta_1(\eta)   & = & \frac{4}{9}\left ( -3 \eta + 
2 \eta^2 + 2(2-\eta)\sqrt{\eta(\eta+1)} \right ), \nonumber \\
\beta_2(\eta)   & = &  1 + 2 \eta  -2 \sqrt{\eta(\eta+1)}, 
\eea
with $\eta := \nu^2/Q^2$. 
Furthermore, $F^p_2(Q^2)$ is the Pauli form factor of the proton, which is
defined in terms of the charge monopole and magnetic dipole
form factors as $F^p_2(Q^2)=(G^p_M(Q^2)-G^p_C(Q^2))/(1+\tau)$
with $F^p_2(0)=\kappa$ being the anomalous part of the proton magnetic
moment. Note that both terms in the integrand of $\delta_1$ diverge  
for $Q^2=0$ but the singularity coming from the second term 
is cancelled by an analogous singularity of the first term
according to the Drell-Hearn, Gerasimov sum rule~\cite{raf71}.

The polarization contribution has been calculated by several
authors~\cite{car08,fau00}. However, only the contribution 
of the magnetic dipole (M1) transition to the $\Delta(1232)$ 
has been studied in some detail. There has been no prior attempt 
to calculate the contribution of the charge quadrupole (C2) transition to 
the $\Delta(1232)$. In order to investigate the effect of the latter 
on the polarization shift, we express the spin structure functions in terms 
of virtual photon absorption cross sections~\cite{dre01} 
\bea
\label{virtual_photoabsorption}
g_1(\nu,Q^2) & = & \frac{m_p \, \nu \, 
(1-Q^2/(2 \, m_p \, \nu))}{8 \pi^2 \alpha \,(1+ Q^2/\nu^2)} \nonumber \\
& & \hspace{-2.1 cm} 
\biggl ( \, \sigma_{1/2}(\nu, Q^2) - \sigma_{3/2}(\nu, Q^2) -
2\, \frac{Q}{\nu} \, \sigma_{LT}^{\prime}(\nu, Q^2)  \biggr ), \nonumber \\
g_2(\nu,Q^2) & = & \frac{m_p \, \nu \, 
(1-Q^2/(2\, m_p\, \nu))}{8 \pi^2 \alpha \, (1+ Q^2/\nu^2)} \nonumber \\
& & \hspace{-2.1 cm} \left ( -\sigma_{1/2}(\nu, Q^2) + \sigma_{3/2}(\nu, Q^2) -
2\, \frac{\nu}{Q}\,  \sigma_{LT}^{\prime}(\nu, Q^2)  \right ).
\eea
Here, $\sigma_{1/2}$ and $\sigma_{3/2}$ are 
the transverse cross sections with photon-nucleon helicity 1/2 and 3/2, 
and $\sigma_{LT}^{\prime}$ is the longitudinal-transverse interference 
cross section.

For excitation energies below 1 GeV, which are most important for hydrogen hfs,
the cross section is dominated by multipole transitions to specific 
nucleon resonances and can be  written in terms of the helicity 
amplitudes $A_{1/2}$, $A_{3/2}$, and $S_{1/2}$ as~\cite{abe98}  
\bea
\label{cross_section_helicity}
\sigma_{1/2} & = & 2 \pi \frac{m_p}{W} \, b \, \vert A_{1/2} \vert^2, 
\nonumber  \\
\sigma_{3/2} & =  & 2 \pi \frac{m_p}{W} \, b \, \vert A_{3/2} \vert^2, 
\nonumber \\
\sigma_{LT}' & = & -\sqrt{2} \pi \frac{m_p}{W} 
\frac{Q}{\vert {\bf q} \vert} \, b \,
S_{1/2}^* A_{1/2}, 
\eea
where $W$ is the total center of mass energy, ${\bf q}$ the $\gamma N$ 
center of mass three-momentum and $b$ the resonance line shape,
which at resonance reduces to $b=2/(\pi \Gamma_{\pi N})$ 
with $\Gamma_{\pi N}$ being the pion decay width of the resonance.
In particular, for $\Delta(1232)$ excitation,
the helicity amplitudes can be expressed via the inelastic $N \to \Delta$ 
transition form factors~\cite{Buc97} shown in 
Fig.~\ref{figure:scattering} (right) 
\bea
\label{helicity_amp_form_factors} 
A_{1/2}(Q^2) & = & -\frac{e}{2 m_p} \,
\sqrt{\frac{\pi}{K_W}} \, \vert {\bf q } \vert \,  \nonumber \\
& & \left ( G_{M1}^{N \to \Delta}(Q^2) 
-\frac{\vert {\bf q} \vert m_p}{\sqrt{6}} G_{E2}^{N \to \Delta}(Q^2) \right ), 
 \nonumber \\
A_{3/2}(Q^2)& = &-\frac{e}{2 m_p} \, \sqrt{\frac{3\pi}{K_W}} \, 
\vert {\bf q } \vert \, \nonumber \\
& & \left (G_{M1}^{N \to \Delta}(Q^2) 
+\frac{\vert {\bf q} \vert m_p}{3 \sqrt{6}} 
G_{E2}^{N \to \Delta}(Q^2)  \right ), 
\nonumber \\
S_{1/2}(Q^2)& = &- e \sqrt{\frac{2\pi}{K_W}} \, \frac{{\bf q}^2}{6} \,
G_{C2}^{N \to \Delta}(Q^2),
\eea
where $K_W=(m_{\Delta}^2-m_N^2)/(2m_{\Delta})$ is the energy transfer 
in the $\gamma N$ center of mass frame at $Q^2=0$.

The contribution of the $N \to \Delta$ charge quadrupole (C2) transition 
form factor to the polarization shift, which comes solely from the
$\sigma_{LT}'$ part in Eq.(\ref{virtual_photoabsorption}) 
can now be evaluated. We obtain with $m_N=m_p$ 
\bea
\label{quadrupole_pol_shift}
\delta_1^{N \to \Delta}(C2) & = & 
\phantom{-} \frac{9}{2}\, \frac{m_N^3}{m_{\Delta}^2} \,\mu_n \, b \, 
\nonumber \\
& & \hspace{-1.2cm} \int_0^{\infty} \! \frac{dQ}{Q} \,
\beta_1\left (\frac{\nu_{\Delta}^2}{Q^2} \right )\,
G_C^n(Q^2) \, \frac{G_M^n(Q^2)}{\mu_n}, \nonumber \\
\delta_2^{N \to \Delta}(C2) & = & 
-6\, \frac{m_N^3}{m_{\Delta}^2} \,
\mu_n \, b \, \nonumber \\
& & \hspace{-1.2cm} 
\int_0^{\infty} \! \frac{dQ}{Q} \,
\beta_2 \left (\frac{\nu_{\Delta}^2}{Q^2} \right )\,
\frac{\nu_{\Delta}^2}{Q^2}\, G_C^n(Q^2) \, \frac{G_M^n(Q^2)}{\mu_n},
\eea 
where $\nu_{\Delta}=(m_{\Delta}^2-m_N^2+Q^2)/(2m_N)$ is 
the resonance energy in the laboratory frame. To derive 
Eq.(\ref{quadrupole_pol_shift}) we have made use of the form factor relations 
of sect.~\ref{subsec:symmetries} and neglected the 
small $E2\times C2$ contribution. Numerically, we obtain with values for 
$\Lambda_M$ and $m$ as in sect.~\ref{sec:zemach}, and 
$\Gamma_{\pi N}=0.12$ GeV the following estimates 
\bea
\label{quadpolshift}
\delta_{1}^{N\to \Delta}(C2) & = & -0.994 \nonumber \\ 
\delta_{2}^{N\to \Delta}(C2) & = & +0.292  \nonumber \\ 
\delta_{pol}^{N\to \Delta}(C2) & = & -0.16\,  {\rm ppm}.  
\eea
The negative sign of the quadrupole polarization shift is indicative
of a prolate (cigar-shaped) intrinsic quadrupole deformation of the proton's
charge distribution. Interestingly, the charge quadrupole $(C2)$ 
polarization contribution in Eq.(\ref{quadpolshift}) is not small compared 
to the magnetic  dipole $(M1)$ polarization 
$\delta_{pol}^{N\to \Delta}(M1)=-0.12$ ppm 
derived earlier~\cite{fau00, ver65}. 
Therefore, it will be important to also calculate the polarization 
shift induced by transverse electric $(E2)$ excitation of the $\Delta(1232)$.
Finally, we also give our result for the $M1$ contribution 
coming from the convergent $g_2$ integral 
$\delta_{2}^{N\to \Delta}(M1)= -0.69$ ppm compared to 
$\delta_{2}= -0.42$ ppm including additional resonances~\cite{fau00}. 

\vspace{0.5 cm}

\section{Summary}
\label{sec:summary}
The nonzeroness of the empirical $p \to \Delta^+(1232)$ quadrupole
transition form factor provides 
evidence that the charge distribution of the proton ground state
deviates from spherical symmetry. Employing SU(6) spin-flavor 
symmetry as a guide we have derived a relation between the 
$p \to \Delta^+$ quadrupole transition and neutron charge form factors. 
It has been shown that this relation agrees with the experimental data
from low to high momentum transfers. On this basis, we have proposed that 
the proton can be assigned an intrinsic charge quadrupole form factor and 
a positive intrinsic quadrupole moment corresponding to a prolate 
(cigar-shaped) distribution of the proton charge. 

We have then investigated how hydrogen ground state hfs is affected by
the proton's non-spherical charge distribution. We have shown that 
the latter is reflected in a {\it positive} deformation contribution 
to the proton Zemach radius, 
where the increment is given by the modulus of the neutron Zemach radius. 
A second consequence of the proton's prolate shape is a 
polarization shift due to the $p \to \Delta^+(1232)$ charge quadrupole 
transition that is quantitatively described by the neutron charge form factor.
We have presented a numerical estimate for this term 
and found that it provides a {\it negative} contribution 
to the polarization shift in atomic hydrogen hfs that exceeds the one coming 
from the previously calculated $p \to \Delta^+(1232)$ magnetic dipole 
transition.

In view of these results it would be interesting to also explicitly 
calculate the polarization shift induced by the transverse electric 
quadrupole form factor $G^{N\to \Delta}_{E2}$. 
Because the experimental accuracy 
of hfs measurements exceeds the theoretical accuracy by several orders of 
magnitude, hydrogen hfs will remain a high precision probe for 
proton structure for many years to come. It is quite possible 
that the interplay between hfs experiment and theory will provide 
independent evidence for the quadrupole deformation of the proton's charge 
distribution. 
Conceivably, hydrogen hfs is also sensitive to higher magnetic 
multipoles, in particular to an intrinsic magnetic octupole 
term in the proton's spatial current distribution. 
We hope to discuss these matters in a future communication.

\end{document}